\documentclass[%
 aps,
superscriptaddress,
 amsmath,amssymb,
 reprint,%
]{revtex4-1}

\usepackage{graphicx}% Include figure files
\usepackage{dcolumn}% Align table columns on decimal point
\usepackage{bm}% bold math
\usepackage[utf8]{inputenc}
\usepackage[T1]{fontenc}
\usepackage{mathptmx}
\usepackage{etoolbox}

\usepackage{hyperref}

\makeatletter
\def\@email#1#2{%
 \endgroup
 \patchcmd{\titleblock@produce}
  {\frontmatter@RRAPformat}
  {\frontmatter@RRAPformat{\produce@RRAP{*#1\href{mailto:#2}{#2}}}\frontmatter@RRAPformat}
  {}{}
}%
\makeatother
\begin{document}

%\preprint{AIP/123-QED}

\title[Emergent spin Hall conductivity in Tantalum-Rhenium alloys]{Emergent spin Hall conductivity in Tantalum-Rhenium alloys}
% Force line breaks with \\

% \altaffiliation[Also at ]{Physics Department, XYZ University.}%Lines break automatically or can be forced with \\

\author{Felix Janus}
\affiliation{ 
New Materials Electronics Group, Department of Electrical Engineering and Information Technology, Technical University of Darmstadt, Merckstr. 25, 64283 Darmstadt, Germany%\\This line break forced with \textbackslash\textbackslash
}%
\author{Jyoti Yadav}%
\affiliation{ 
New Materials Electronics Group, Department of Electrical Engineering and Information Technology, Technical University of Darmstadt, Merckstr. 25, 64283 Darmstadt, Germany%\\This line break forced with \textbackslash\textbackslash
}%
\affiliation{ 
Spin Dynamics Lab, Department of Physics, Indian Institute of Technology Delhi, New Delhi, Delhi 110016, India%\\This line break forced with \textbackslash\textbackslash
}%
\author{Nicolas Beermann}%
\author{Wentao Zhang}
\author{Hassan A. Hafez}
\author{Dmitry Turchinovich}
\affiliation{ 
Fakultät für Physik, Universität Bielefeld, Universitätsstraße 25, 33501 Bielefeld, Germany%\\This line break forced with \textbackslash\textbackslash
}
\author{Sascha Preu}
\affiliation{ 
THz Devices and Systems, Department of Electrical Engineering and Information Technology, Technical University of Darmstadt, Merckstr. 25, 64283 Darmstadt, Germany%\\This line break forced with \textbackslash\textbackslash
}
\author{Markus Meinert}
 \email{markus.meinert@tu-darmstadt.de}
\affiliation{ 
New Materials Electronics Group, Department of Electrical Engineering and Information Technology, Technical University of Darmstadt, Merckstr. 25, 64283 Darmstadt, Germany%\\This line break forced with \textbackslash\textbackslash
}
%\author{C. Author}
% \homepage{http://www.Second.institution.edu/~Charlie.Author.}
%\affiliation{%
%Second institution and/or address%\\This line break forced% with \\
%}%

\date{\today}% It is always \today, today,
             %  but any date may be explicitly specified

\begin{abstract}
We investigate the spin Hall conductivity (SHC) of a composition series of a Ta-Re bcc solid solution. At approximately 60\,at.\% Ta the Ta-Re alloy features an SHC similar to bcc-W, while both endpoints of the compositional series have rather moderate SHC. The intermediate stoichiometries exhibit a substantial enhancement due to Fermi level tuning through the same band structure feature which gives bcc-W and $\beta$-W its large SHC. We provide experimental evidence for the enhancement of the SHC in the alloy via THz emission upon ultrafast laser excitation of Ta-Re/CoFeB bilayers. We demonstrate that a rigid band model derived from bcc-W and Fermi level tuning describes the experimental data with a similar accuracy as coherent potential approximation alloy calculations of the SHC.

\end{abstract}

\maketitle

\section{\label{sec:intro}Introduction}
The spin Hall effect (SHE) in metals converts a charge current density $j_\mathrm{c}$ into a transverse spin current density $j_\mathrm{s}$, which enables the manipulation of the magnetization of a magnetic thin film in thin multilayer structures.\cite{Jungwirth2012, Hoffmann2013} The SHE is usually quantified in terms of the spin Hall angle (SHA) $\theta_\mathrm{SH} = (2e/\hbar) j_\mathrm{s} / j_\mathrm{c}$. In the moderately dirty regime of typical metallic thin films (with resistivities $\rho$ ranging from 20\,$\mu\Omega$cm to 300\,$\mu\Omega$cm), the intrinsic mechanism is dominant and the SHA can be written as $\theta_\mathrm{SH} = \sigma_\mathrm{SH} \rho$ in isotropic metals, with the isotropic spin Hall conductivity (SHC) $\sigma_\mathrm{SH}$. The direct dependence on the resistivity enables tuning of the SHA by either alloying or deliberate introduction of defects via deposition process parameters, while maintaining the crystal structure and thus the SHC.\cite{Sagasta2016} The elements with the largest SHC are Tungsten (W) and Platinum (Pt). This originates from the coincidence of the Fermi energy with negative or positive peaks in the Fermi energy-dependent SHC, respectively.\cite{Tanaka2008, Qiao2018, McHugh2020} For Pt, the Fermi energy coincides almost perfectly with the maximum, whereas W has its peak slightly below the Fermi energy. Thus, the SHA in Pt can only be tuned by resistivity enhancement.\cite{Obstbaum2016, McHugh2024} In contrast, W allows for Fermi level tuning of the SHC and simultaneously for resistivity tuning by alloying.\cite{Sui2017, Kim2020, Coester2021} Another well-known element with a large SHC is Tantalum (Ta). Both W and Ta have metastable phases with complex crystal structures: $\beta$-W forms the A15 structure with an eight-atomic primitive cubic cell,\cite{Chen2018} whereas $\beta$-Ta has the $\beta$-U structure with a 30 atom tetragonal unit cell.\cite{Jiang2003} Both structures allow for structural disorder, which leads to large resistivity values between 100\,$\mu\Omega$cm and 200\,$\mu\Omega$cm. SHA and SHC tuning were demonstrated in multiple alloy systems based on Ta, W, or Pt.\cite{Qu2018, Fritz2018, Meinert2020}

To decouple the occurence of the large SHC from the element W and unambigously show the Fermi-level tuning through the SHC, we carried out a combined experimental and theoretical investigation of a solid solution of Tantalum and Rhenium (Re). In the periodic table of the elements, Ta and Re are positioned as direct neighbors to W. A solid solution of the two elements should resemble the SHC of W at approximately equiatomic composition. The binary phase diagram of the Ta-Re system indicates formation of a bcc solid solution to at least 40\% of Re in Ta.\cite{Liu2000, Berne2001} At higher Re content, more complex phases (so-called $\sigma$- and $\chi$-phases) appear in the phase diagram. Ta is found to be almost insoluble in Re. Rhenium crystallizes in the hexagonal system with a nearly close-packed structure ($c/a \approx 1.613$ is very close to the ideal value of 1.63). It is a refractory metal that finds its use cases in the enhancement of mechanical properties in Nickel-based superalloys.\cite{Wee2020} Only one study reports about the spin Hall effect of Re and ReO$_x$ and found a very small spin Hall angle in crystallized hcp-Re.\cite{Karube2020}

This paper is organized as follows: we first outline our theoretical and experimental methods. Next, we present theoretical results on the solid solution of the Ta-Re system with several simple structural models, including spin Hall conductivities and formation energy calculations. Afterwards, we present our results on THz emission measurements from thin film samples of the Ta-Re alloys and compare those with our theoretical predictions.

\section{Methods}

\subsection{Theory}

Density functional theory calculations for a rigid-band model of the solid solution were performed with the elk full-potential linearized augmented planewave code\cite{elk} with the Perdew-Burke-Ernzerhof generalized gradient approximation functional. All convergence parameters were chosen for high accuracy and the k-point mesh was set to $15 \times 15 \times 15$ points for the self-consistency cycle. Density of states (DOS) calculations were then perfomed on interpolated $200^3$ meshes. The energy-dependent spin Hall conductivity was calculated with the Wannier90 code on $50 \times 50 \times 50$ k-point meshes with auxiliary $4 \times 4 \times 4$ fine meshes in regions with high contributions to the SHC.\cite{Qiao2018} Lattice constants for the hypothetical structures of the elements were taken from the Materials Project database.\cite{Jain2013}

The spin Hall conductivities of the solid solution model were calculated in a fully relativistic multiple-scattering Green function framework using the Kubo-Bastin formalism \cite{Lowitzer2011}. Intrinsic and extrinsic contributions to the spin Hall conductivity are treated on equal footing. Furthermore, chemical alloying as well as temperature-induced disorder are accounted for within the coherent potential approximation (CPA) or the alloy-analogy model (AAM)\cite{Ebert2015}, respectively. The formalism is implemented in the Munich Spin-Polarized Relativistic Korringa-Kohn-Rostoker (SPR-KKR) program package \cite{Ebert2011}. The Green function was expanded up to $\ell_\mathrm{max} = 3$ and the Fermi energy was corrected with Lloyd's formula. The atomic sphere approximation (ASA) was used throughout. For the evaluation of the Kubo-Bastin formula 32 points were used for the energy integration. $168 \times 168 \times 168$ k-points ($125 \times 125 \times 125$ for $\beta$-W, $165 \times 165 \times 88$ for hcp-Re) in the full Brillouin zone were used to ensure an accurate evaluation of the Brillouin zone integrals for the Fermi surface term. The atomic volumes and the Debye temperatures are interpolated between the elemental values according to Vegard's rule. The spin Hall conductivities were calculated at 300\,K in all cases. We note that we use the convention of displaying the spin Hall conductivity in units of $\frac{\hbar}{2e}$ S/m throughout the paper. In addition, alloy formation energies per atom $\Delta E$ were calculated as
\begin{align}
\Delta E_\alpha (x) = &E_\alpha (\mathrm{Ta}_{1-x} \mathrm{Re}_{x}) \nonumber\\&- \left[ (1-x) E_\mathrm{bcc}(\mathrm{Ta}) + x E_\mathrm{hcp}(\mathrm{Re})\right],
\end{align}
where $\alpha = \mathrm{fcc}, \mathrm{bcc}, \mathrm{hcp}$ and $E_\alpha$ represents the elemental total energies per atom.

\subsection{Experiment}
Thin films multilayer structures of fused silica / Ta$_{1-x}$Re$_x$ 5.5\,nm / CoFeB 2.5\,nm / Ta 0.5\,nm / TaO$_x$ 1\,nm were deposited by magnetron co-sputtering. Here, CoFeB denotes the composition Co$_{40}$Fe$_{40}$B$_{20}$. For comparison, similar samples with W instead of the Ta-Re alloy were prepared by the same method. An additional stack without the heavy metal film was made to determine the shunt resistance in the other films. The substrates were held at room temperature, the deposition pressure was $2 \times 10^{-3}$\,mbar, the base pressure was below $5 \times 10^{-8}$\,mbar. The deposition rates of all films were typically around 0.1\,nm/s. The substrates were rotated at 30\,rpm to avoid the formation of uniaxial magnetic anisotropy. Growth rates were determined by x-ray reflectivity.

The film resistivities were measured with a standard four-point probing technique. Because the probe spacing is 3\,mm while the substrate size is $10 \times 10$\,mm$^2$ (with a film area of only $8 \times 8$\,mm$^2$), the usual analytical formula for the resistivity $\rho = (\pi / \mathrm{ln}2) R_{4w} t$, with the four-wire resistance R$_{4w}$ and the metal film thickness $t$ is not applicable here. We determined the prefactor of this formula by finite-element calculations and obtain $\rho = 2.3 R_{4w} t$ for central, diagonal placement of the four points across the square sample.

X-ray diffraction measurements were performed with a Rigaku rotating anode diffractometer with a Cu anode and a Ge monochromator.

Broadband ferromagnetic resonance spectroscopy with the OpenFMR system\cite{Meinert2024} was used to determine the effective magnetization and the Gilbert damping parameter of all samples.

The spin Hall quantities were determined qualitatively by THz emission spectroscopy (TES).\cite{Seifert2016, Meinert2020, Papaioannou2021} In TES, a heavy metal (HM) $/$ ferromagnet (FM) bilayer is excited by a femtosecond laser pulse, thereby inducing ultrafast spin transport from the FM into the HM layer through an ultrafast version of the spin-dependent Seebeck effect.\cite{Seifert2016, Kampfrath2013, Seifert2017, Alekhin2017} In the HM, the laser-driven longitudinal spin current is converted into a transverse charge current by the inverse SHE. The resulting sub-picosecond charge current\cite{Seifert2018} gives rise to the emission of electromagnetic radiation at THz frequencies.\cite{Seifert2016} We initially used an amplified Ti:sapphire laser system with a pulse width of 100\,fs centered at a wavelength of 800\,nm, a repetition rate of 1000\,Hz (chopped at 500\,Hz), and a power of 250\,mW. The samples were exposed to a magnetic field of 60\,mT to align the magnetization. The measurement of the THz electric field was performed with electro-optic sampling (EOS) in a 500\,$\mu$m ZnTe crystal and lock-in detection on a balanced detector. For finer sampling of the stoichiometry series, we additionally used a modified time-domain THz spectroscopy system from Menlo Systems employing a fiber laser with a wavelength of 1550\,nm, a pulse width of 90\,fs, a power of 300\,mW, and a repetition rate of 100\,MHz.\cite{Nandi2019} The detection was done with a photoconductive switch (PCS) and a transimpedance amplifier with 5000 delay sweeps per measurement. A second-order low-pass Butterworth filter with a cutoff frequency of 2\,THz was applied to reduce the noise in the acquired waveforms of the THz spectrometer. To facilitate a quick sample alignment, nearly parallel laser beams were used, which allow for a sample changing without realignment.

\begin{figure}[t]
\includegraphics[width=8.6cm]{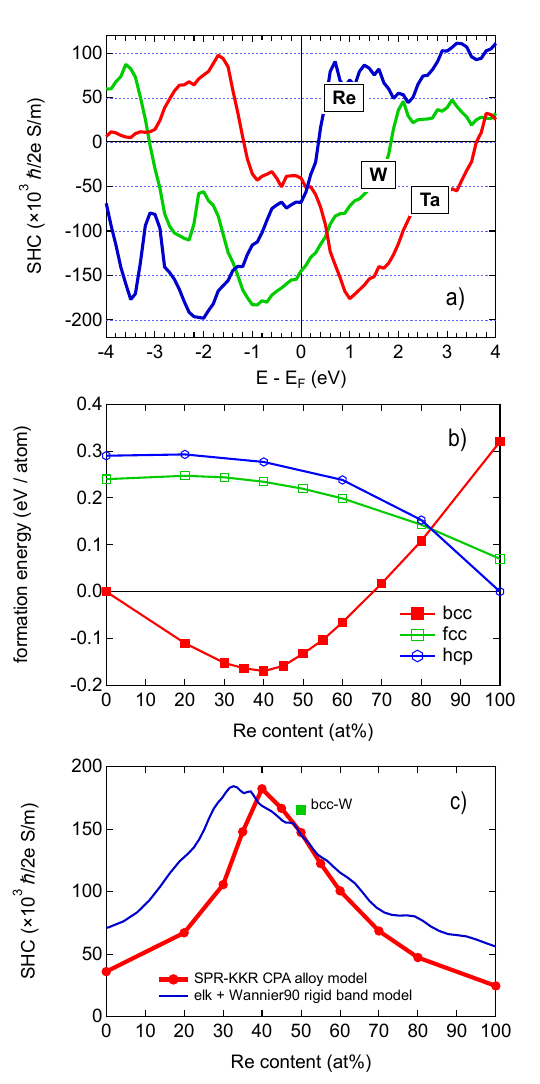}
\caption{\label{fig:Ta-Re_theory}
a) Fermi-energy dependent spin Hall conductivities of bcc-Ta, bcc-W, and bcc-Re. b) Formation energies of the Ta-Re solid solution with bcc, fcc, and hcp structures. c) Spin Hall conductivity of the solid solution within two models: CPA alloy model (calculated with SPR-KKR), and rigid band model (calculated with elk and Wannier90). The green square represents the SHC of bcc-W calculated with SPR-KKR.}
\end{figure}

The THz emission amplitudes are modeled as a function of THz frequency $\omega/2\pi$ as\cite{Seifert2016,Yang2024}
\begin{equation}\label{eq:THz}
E_\mathrm{THz}(\omega) = {A\cdot B(\omega) \cdot \lambda_\mathrm{HM}}\cdot\tanh\left(\frac{t_\mathrm{HM}}{2\lambda_\mathrm{HM}}\right)\cdot\theta_\mathrm{SH}\cdot Z(\omega).
\end{equation}
Here, $A$ is the pump-light absorptance, while the factor $B$ captures the photon-to-spin-current conversion efficiency and the detector response function. $B$~is assumed to be independent of the alloy composition in our experiment, thereby neglecting possible variations of the spin-current strength due to, e.g., variations of the interface quality for different stoichiometries. We omit its film thickness dependence due to the constant thickness of our films. The spatial shape of the spin current in the HM layer is captured by the spin-current relaxation length $\lambda_\mathrm{HM}$, which becomes comparable to the mean free path length at THz frequencies.

The charge-current-to-electric-field conversion is described by the multilayer impedance
\begin{equation}\label{eq:impedance}
Z(\omega) = \frac{Z_0}{n_1(\omega) + n_2(\omega) + Z_0 \int_0^d \mathrm{d}z \sigma_{xx}(z, \omega)}
\end{equation}
where $n_1(\omega)$ and $n_2(\omega)$ are the refractive indices of air and the substrate, respectively, $Z_0 = 377\,\Omega$, and $\sigma_{xx}(z, \omega)$ is the in-plane conductivity of the material at depth $z$. For simplicity, we take $\sigma_{xx}$ as constant across the film thickness, such that the integral turns into a sum of the reciprocal resistances per square. We further ignore the frequency dependence, because the frequencies used here are well below the Drude frequency of the material.

The SHA relative to a reference sample can be obtained for all alloy stoichiometries when $\lambda_\mathrm{HM}$ is known. Here, we take $\lambda_\mathrm{HM}$ as the electron mean free path
$\lambda_\mathrm{MF}$ and use $\lambda_\mathrm{MF}/\sigma_{zz} = \kappa = 8.6\times 10^{-16}~\Omega~\mathrm{m}^2$ where $\sigma_{zz}$ is the electrical conductivity of the HM perpendicular to the film plane.\cite{Gall2016} $\kappa$ is a parameter that is derived from the properties of the Fermi surface and the electron scattering of the material. For transition metals it is typically in the range of 3 to  $9\times 10^{-16}~\Omega~\mathrm{m}^2$.\cite{Gall2016, Dutta2017} Measuring $\sigma_{zz}$ is impractical, so we employ the approximation $\sigma_{zz} \approx \sigma_{xx}$, which is justified in materials where the film thickness is much larger than the mean free path. Fine-tuning the model to stoichiometry-dependent electron mean free paths does not substantially change any conclusions because our film thickness is larger than $4\lambda_\mathrm{MF}$, thus $\tanh\left(\frac{t_\mathrm{HM}}{2\lambda_\mathrm{HM}}\right) \approx 1$ for all compositions. Under these assumptions, we can write $\lambda_\mathrm{HM} \theta_\mathrm{SH} \approx \kappa\sigma_{xx} \sigma_\mathrm{SH}/\sigma_{xx} = \kappa \sigma_\mathrm{SH}$ and Eq. \ref{eq:THz} becomes
\begin{equation}
E_\mathrm{THz}(\omega) \approx A \cdot B(\omega) \cdot \kappa\cdot\sigma_\mathrm{SH}\cdot Z(\omega).
\end{equation}

Note that $\theta_\mathrm{SH}$ of Eq. \ref{eq:THz} is an effective SHA which, in addition to spin-to-charge-current conversion in the HM layer, contains such conversion also in the FM layer and at the FM/HM interface. For the evaluation of Eq. \ref{eq:THz} we drop the frequency dependendence and use the peak-to-peak difference of the detected electric field $E(t)$ instead of the spectral amplitude $E(\omega)$. As the temporal shape of the THz waveforms is found to not be compositon-dependent, it is sufficient to evaluate the peak THz field amplitude in the time domain instead of the frequency domain as a relative measure for the emitted field and thus the SHC.

\section{Results}

\subsection{Calculations}

We begin the discussion of results by investigating the Fermi-level dependences of the spin Hall conductivites of bcc-Ta, bcc-W, and bcc-Re, as calculated with Wannier90. The scans are shown in Figure \ref{fig:Ta-Re_theory} a). Our results are in good agreement with previous calculations of Ta and W with a similar approach.\cite{Sui2017, Qiao2018, Derunova2019,McHugh2020} The general shape of the three curves is very similar with a pronounced negative peak close to the Fermi energy. In bcc-W, the Fermi energy is very close to the extremum, whereas the lower (higher) electron concentration in Ta (Re) shift the Fermi energy below (above) the extremum. Both Ta and Re are predicted to have a similar spin Hall conductivity around $-50\,000$\,$\frac{\hbar}{2e}$ S/m, whereas for W we find approximately $-150\,000$\,$\frac{\hbar}{2e}$ S/m. The structure of the curves indicates that a rigid-band model provides a reasonable understanding of the SHC in the alloy system. Thus, a bcc solid solution of Ta and Re at approximately the equiatomic composition should have an SHC similar to bcc-W.

To assess the feasibility of the solid-solution formation, we calculated the formation energies of bcc, fcc, and hcp solid solutions in the Ta-Re binary system (see Figure \ref{fig:Ta-Re_theory} b)). Our calculations reproduce the correct ground states, bcc for Ta and hcp for Re. According to these calculations, a bcc solid solution may exist up to nearly 70\% Re at low temperature. In contrast, the fcc and hcp solid solutions have positive formation energy at all compositions and might only occur as entropically stabilized high-temperature phases. Our calculations are thus in line with the experimental result that Ta is essentially not soluble in Re at low temperature, whereas a Ta-rich bcc solid solution exists.

We proceed by calculating the SHC of the bcc solid solution of Ta-Re in a rigid-band model derived from the Fermi-energy scan of the SHC in bcc-W. We obtain the number of states from integrating the total density of states. Interpolating and inverting yields the Fermi energy shift as a function of the valence electron concentration. This model predicts somewhat higher SHC for bcc-Ta (compared to the direct calculation of Ta). This can be easily explained by the larger lattice constant of bcc-Ta, which gives a narrower bandwidth and thereby a sharper SHC peak, which is not correctly accounted for by the rigid-band model. This model predicts an extremum of the SHC between 30\% to 40\% Re content in the bcc alloy with a peak value of $-184\,300$\,$\frac{\hbar}{2e}$ S/m. This is about 4.5 times as large as the calculated SHC of bcc-Ta ($-40\,600$\,$\frac{\hbar}{2e}$ S/m).

The CPA calculation (Fig. \ref{fig:Ta-Re_theory} c)) of the solid solution resembles the general trend of the rigid-band model, but is more peaked. It reproduces the SHC of Ta accurately, also the peak value agrees perfectly with the rigid-band model. However, the SHC of bcc-Re is grossly underestimated compared to the Wannier90 calculation. Further, we find that the peak position is shifted to higher Re concentration (higher valence electron concentration or higher Fermi energy). For comparison, we also calculated the SHC of bcc-W within the same framework and find a slightly higher SHC as compared to the Wannier90 calculations. The deviation between the methods may be due to slightly different placement of the Fermi energy, or the inclusion of temperature (which generally reduces the SHC) or inclusion of vertex corrections. For Re, a slightly too high Fermi energy will lead to an underestimation of the SHC with respect to the Wannier90 calculation. Generally, we consider the agreement between the methods and previous calculations as excellent. For completeness, we also considered the spin Hall conductivities of hcp-Re and $\beta$-W. For hcp-Re, we find three independent components: $\sigma_{yx}^z \approx -15\,100\,\frac{\hbar}{2e}$ S/m, $\sigma_{zy}^x \approx -161\,800\,\frac{\hbar}{2e}$ S/m, and $\sigma_{xz}^y \approx -128\,900\,\frac{\hbar}{2e}$ S/m. Here, the $y$ direction is along the hexagonal axis in the basal plane and $z$ is perpendicular to the basal plane. For $\beta$-W we obtain $-488\,000$\,$\frac{\hbar}{2e}$ S/m, with a room-temperature resistivity of 26\,$\mu\Omega$cm. It is thus important to note that $\beta$-W does not intrinsically have a high resistivity; instead, the crystal structure occurs when W is contaminated with impurities, such as oxygen, which then lead to the high resistivity. Light impurities will reduce the spin Hall conductivity and bring it in line with typical experimental findings.\cite{McHugh2020}

With this theoretical guidance in mind, we manufactured spintronic THz emitters with the Ta-Re alloys on fused silica substrates for a standard transmission geometry with laser illumination from the substrate side as described above. 

\subsection{Experiment}

\begin{figure}[t]
\includegraphics[width=8.6cm]{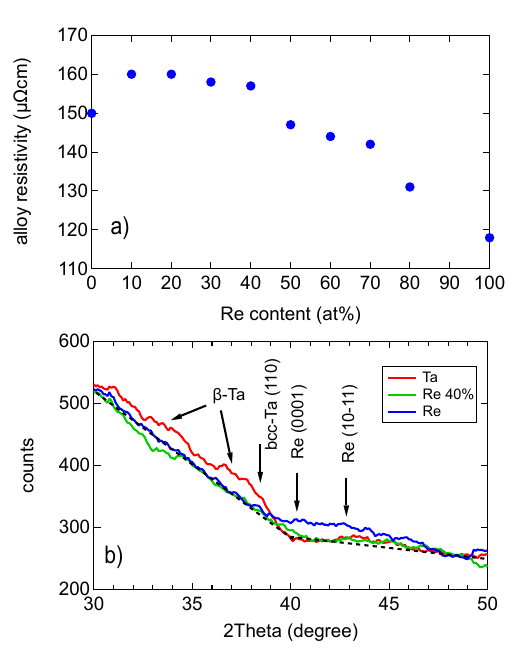}
\caption{\label{fig:Ta-Re_resistivity_XRD}
a)  Ta-Re alloy resistivity. b) XRD measurements with Cu $K_\alpha$ radiation on three Ta-Re film samples. Vertical arrows indicate expected peak positions. The dashed line indicates the background from the fused silica substrate.}
\end{figure}

As expected for sputter-deposited films, we find a high-resistivity phase of Ta with $\rho = 150\,\mu\Omega$cm, which is usually identified as $\beta$-Ta. By adding Re, the resistivity is slightly increased and monotonously decreases with Re addition beyond 10\,\% (Fig. \ref{fig:Ta-Re_resistivity_XRD} a)). The Re film has $\rho = 118\,\mu\Omega$cm, indicating the formation of a high-resistivity phase of Re instead of a crystalline hcp-Re film. We performed x-ray diffraction measurements on three films: Ta, Re, and Ta$_{0.6}$Re$_{0.4}$. The fused silica substrate yields a large, characteristic background. This complicates weak-peak identification due to the Poisson noise of the single-photon counting being proportional to the square-root of counts. We thus apply a simple boxcar filter with a 15-point window (a 1.5$^\circ$ range) to the data to reconstruct the underlying spectrum via a local maximum likelihood estimation. The datasets are shown in Fig. \ref{fig:Ta-Re_resistivity_XRD} b). By directly comparing the three diffraction measurements, we identify the typical $\beta$-Ta reflections below 40$^\circ$ on top of the sloped background and the absence of an isolated bcc-Ta peak. The Re film exhibits a broad reflection between 39$^\circ$ and 47$^\circ$, with its center at approximately 43$^\circ$. This coincides with the hcp-Re $(10\bar{1}1)$ reflection and rules out the formation of a metastable fcc phase, which would have its $(111)$ reflection at the same position as hcp-Re $(0001)$. This result also indicates an unconventional growth orientation: conventionally one finds (0001) texture for hcp films, whereas our result indicates a $(10\bar{1}1)$ texture. We estimate the full-width-at-half-maximum as $\Delta 2\theta \approx 4.6^\circ$; according to the Scherrer formula, this corresponds to a crystallite size of 1.9\,nm. This is smaller than the film thickness of 5.5\,nm and indicates the formation of a nanocrystalline state. Therefore, the resistivity of the Re film remains high compared to crystalline bulk material $(\rho = 19.3\,\mu\Omega$cm at room temperature). Remarkably, the alloy exhibits no identifiable structures and is thus assessed as an amorphous alloy. This may explain the slight increase of the resistivity compared to the Ta film, which is nanocrystalline. With $\kappa = 8.6\times 10^{-16}~\Omega~\mathrm{m}^2$ the mean free path can be estimated as approximately $\lambda_\mathrm{MF} \approx 0.55$\,nm for the approximately equiatomic alloys. Thus, our model assumption for the THz emission of $\sigma_{xx} = \sigma_{zz}$ is justified.

Broadband FMR measurements demonstrated an effective magnetization of $M_\mathrm{eff} \approx 875 \pm 25$\,kA/m without any trend across the stoichiometry series. In contrast, the Gilbert damping parameter $\alpha$ increased from 0.013 (Ta) to 0.02 (Re) nearly linearly. A slightly reduced value $M_\mathrm{eff}$ as compared to the bulk value of $M_\mathrm{eff} \approx 1050$\,kA/m (measured for a 25\,nm CoFeB film) is expected and is due to reduced Curie temperature of the thin film and/or due to the presence of an interface anisotropy. The Gilbert damping is significantly larger than the bulk value for our CoFeB of 0.0055. The increase of the Gilbert damping across the alloy series may be interpreted in terms of an enhanced spin-mixing conductance at the interface, and enhanced spin-orbit coupling in Re as compared to Ta (which leads to stronger spin scattering), or modifications of the interface quality.

\begin{figure}[t]
\includegraphics[width=8.6cm]{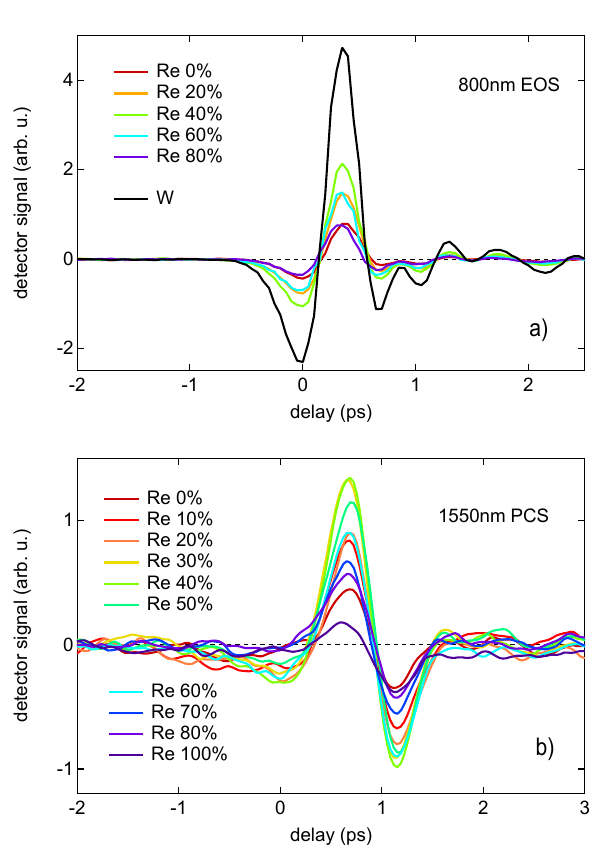}
\caption{\label{fig:Ta-Re_THz_waveforms}a) THz waveforms obtained with the 800nm EOS setup. In addition to select Ta-Re alloys we show the waveform of our $\beta$-W reference sample. b) THz waveforms obtained with the 1550nm PCS setup. Due to worse signal-to-noise ratio, the data were low-pass filtered with a cutoff frequency of 2\,THz, which does not change the principal shape of the waveform but reduces noise substantially. 
}
\end{figure}

\begin{figure}[th!]
\includegraphics[width=8.6cm]{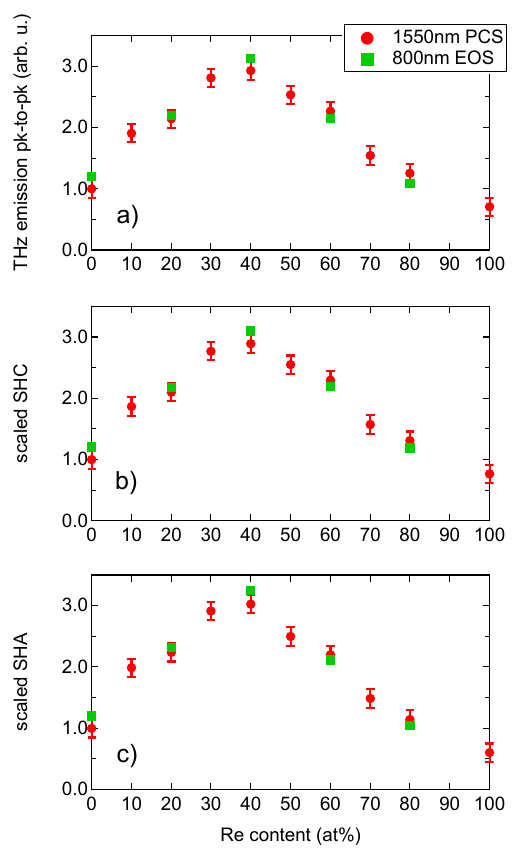}
\caption{\label{fig:Ta-Re_THz_emission}
a) THz emission peak-to-peak reading, normalized to the value of Ta. b) and c) Spin Hall conductivity and spin Hall angle obtained via Eq. \ref{eq:THz}.}
\end{figure}

The THz emission measurements (Fig. \ref{fig:Ta-Re_THz_waveforms} a) and b)) show the typical waveforms\cite{Seifert2016} with a substantial variation in the electric field amplitude across the alloy series. We observe that the emission amplitude increases by approximately 2.7 times with addition of Re. However, the W reference film ($\rho = 141\,\mu\Omega$cm) has even larger emission amplitude with nearly seven times the amplitude of Ta. The high resistivity of the W film indicates the formation of the $\beta$-phase, for which a huge SHC is expected. Notably, the observed trend of Ta $\rightarrow$ Ta$_{60}$Re$_{40}$ $\rightarrow$ $\beta$-W is perfectly in line with the calculated spin Hall conductivities discussed above.  In Fig. \ref{fig:Ta-Re_THz_emission} a) - c) we show the THz peak-to-peak amplitudes, and the extracted SHC and SHA according to Eq. \ref{eq:THz}. The data are scaled to the Ta as the reference. Because the resistivity does not change much across the alloy series, the emission amplitude, the scaled SHC, and the scaled SHA look very similar, with similar ratios between the peak and the two endpoints of the series. Furthermore, we observe that the agreement between the measurements with the different laser systems is very good, when the peak-to-peak amplitude data from the 800\,nm EOS system is scaled as a "best fit" to the data from the 1550\,nm PCS system within the error bars.

\begin{figure}[t]
\includegraphics[width=8.6cm]{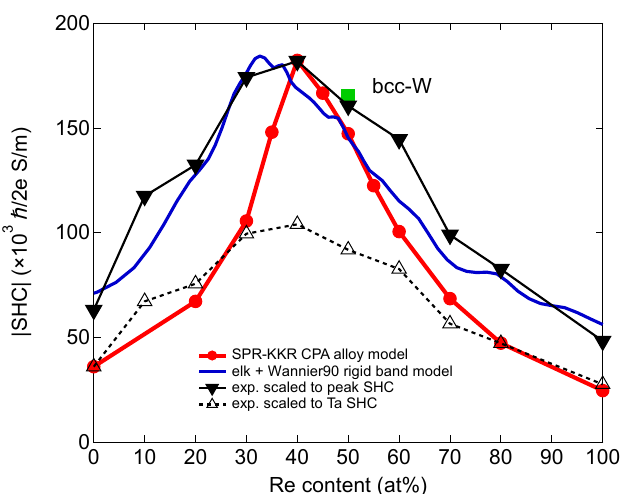}
\caption{\label{fig:Ta-Re_comparison} Comparison of experimental data (black) and theoretical results (red, blue, green). The theoretical data are identical to Fig. \ref{fig:Ta-Re_theory} c). Experimental data are scaled to the theoretical value of bcc-Ta (dashed black line and open triangles) and to the theoretical peak value (solid line and filled triangles). 
}
\end{figure}

In Fig. \ref{fig:Ta-Re_comparison} we compare the extracted (scaled) spin Hall conductivities to the theoretical data presented in Fig. \ref{fig:Ta-Re_theory} c). We apply two scalings to the experimental data: first we scale to the SHC of bcc-Ta (dashed line, open triangles), and second, we scale to match the peak values of the models (solid line, filled triangles).

Over the full range of composition, the peak-scaled measured data are remarkably similar to the rigid-band model. In contrast, the data scaled to the bcc-Ta SHC show a substantially too small peak as compared to the theoretical data. Within the scaling to the theoretical peak, the pure Ta emitter has an SHC of $-63\,000$\,$\frac{\hbar}{2e}$ S/m, which yields $\theta_\mathrm{SH} = -0.0945$ with $\rho = 150\,\mu\Omega$cm, close to typical experimental results but still somewhat smaller.\cite{Neumann2018, Xie2023} As the SHC of the sputter-deposited $\beta$-Ta films is well-known to exceed the SHC of bcc-Ta, it is not surprising that the Ta-rich portion of the alloy series has a flatter stoichiometry dependence than suggested by the CPA alloy model because of the higher starting point. Even though our alloy appears to be amorphous, its local coordination should resemble a bcc structure, which is the only structure with a negative formation energy. As was shown previously on W-Hf alloys, CPA calculations can reproduce experimental data on amorphous alloys surprisingly well.\cite{Fritz2018} The theoretical peak value of $-183\,000$\,$\frac{\hbar}{2e}$ S/m implies $\theta_\mathrm{SH} = -0.29$ with the experimental resistivity of $\rho = 158\,\mu\Omega$cm for Ta$_{60}$Re$_{40}$. Comparison with the W reference sample indicates that Ta$_{60}$Re$_{40}$ should have an SHC roughly 45\% of $\beta$-W. With the calculated value of $-488\,000$\,$\frac{\hbar}{2e}$ S/m for $\beta$-W, this implies  $\sigma_\mathrm{SH} \approx -222\,000$\,$\frac{\hbar}{2e}$ S/m for Ta$_{60}$Re$_{40}$. This estimate is close to the calculated peak value and gives some reassurance that the peak-scaled data is self-consistent. The Re-rich portion of the alloy series behaves again quite similar to the rigid-band model of bcc-W. The Re film has lower SHC than the Ta film. The hcp structure has a high calculated spin Hall conductivity (except for spin alignment along the c-axis), as well as the fcc structure (for which we calculate $\sigma_\mathrm{SH} = -136\,600\,\frac{\hbar}{2e}$ S/m). Our XRD results rule out fcc-Re and indicate an unusual $(10\bar{1}1)$ texture of hcp-Re, which means that the $c$-axis is at $\alpha \approx 28^\circ$ with the surface. To transform the calculated SHC tensor components ($\sigma_{yx}^z \approx -15\,100\,\frac{\hbar}{2e}$ S/m, $\sigma_{zy}^x \approx -161\,800\,\frac{\hbar}{2e}$ S/m, and $\sigma_{xz}^y \approx -128\,900\,\frac{\hbar}{2e}$ S/m) into the coordinate system of the thin film, we apply a tensor rotation $\sigma_{\alpha, \beta}^\gamma = R_{\alpha A} R_{\beta B} R_{\gamma C} \sigma_{A, B}^C$ (with the Einstein summation convention and the proper rotation matrix $R$) about the crystallographic $b$-axis and consecutively determine the SHC components for spin in the plane and determine the average for the isotropic crystallite orientation in the plane. This yields $\sigma_{\left<zy\right>}^x =- 91\,800\,\frac{\hbar}{2e}$ S/m and is consistent with our observation that the SHC of Re is similar to that of Ta. However, the experimental value extracted from the peak-scaling model of Fig. \ref{fig:Ta-Re_comparison} is still smaller than the calculated value.

\section{Conclusion}
We present experimental evidence for Fermi level tuning of the spin Hall conductivity in an amorphous Ta-Re alloy system, where the THz emission amplitude at 40\% Re is 2.9 times higher as compared to a pure Ta film. We interpret this result with the standard THz emission model as a massive increase in the spin Hall conductivity. By comparison with theoretical results for a rigid-band model and CPA alloy calculations, we identify the observed peak of the spin Hall conductivity as Fermi-level shifting through a peak in the energy-dependent spin Hall conductivity. Because of the amorphous nature of the alloy, the resistivity is very high. Together with the large spin Hall conductivity inferred from the THz emission measurements, this alloy is a promising candidate for advanced spin-Hall-effect based spintronic devices: the results suggest a spin Hall angle of $\theta_\mathrm{SH} = -0.29$ for Ta$_{60}$Re$_{40}$.  Future studies should focus on thicker films with better crystallinity and even epitaxial layers where the substrate can impose the bcc or hcp structure to create a testbed for more detailed studies on Fermi energy tuned spin Hall conductivity. For practical applications, one has to study the annealing temperature dependence of the Ta-Re alloy and assess the phase stability upon heat treatment. Because $\beta$-W tends to recrystallize into its low-resistivity bcc form upon heat exposure,\cite{Neumann2016} an amorphous Ta-Re alloy may be a suitable replacement in high-power applications, where the phase stability of $\beta$-W is a major concern.

\begin{acknowledgments}
The Darmstadt groups acknowledge funding by the Deutsche Forschungsgemeinschaft (DFG) under Project Number 513154775, and under the Major Instrumentation Programme Project Numbers 468939474, and 511340083. The group in Bielefeld acknowledges the financial support from the European Union’s Horizon 2020 research and innovation program (Grant Agreement No.964735 EXTREME-IR), Deutsche Forschungsgemeinschaft (DFG) within Project No. 468501411-SPP2314 INTEGRATECH, Bundesministerium für Bildung und Forschung (BMBF) within Project No. 05K2022 PBA Tera-EXPOSE, and Bielefelder Nachwuchsfond. Funding from the joint DFG project under Project Number 518575758 (HIGH-SPINTERA) is acknowledged by the Bielefeld and Darmstadt groups.

\end{acknowledgments}

\section*{Data Availability Statement}

The data that support the findings of this study are available upon reasonable request from the corresponding author.

\section*{References}

\bibliography{cite} % Produces the bibliography via BibTeX.

\end{document}